\documentclass[pre,preprint,amsmath,showpacs,amssymb]{revtex4}
\usepackage{graphicx,psfrag}

\begin{document}
\title{Phase effects on synchronization by dynamical relaying in delay-coupled systems}
\author{Chitra R N }
\email{rchitra@cusat.ac.in}
\author{V C Kuriakose}
\email{vck@cusat.ac.in} %
\affiliation{Department of Physics, Cochin University of Science
and
Technology, Kochi, 682022} %
\begin{abstract}
Synchronization in an array of  mutually coupled systems with a
finite time-delay in coupling is studied using  Josephson
junction as a model system. The sum of the transverse Lyapunov
exponents is  evaluated as a function of the parameters by
linearizing the equation about the synchronization manifold. The
dependence of synchronization on damping parameter, coupling
constant and time-delay is studied numerically. The change in
the dynamics of the system due to time-delay and phase
difference between the applied fields is studied. The case where
a small frequency detuning between the applied fields is also
discussed.
\end{abstract}

\maketitle

\textbf{Time-delayed systems are  interesting because the
dimension of  chaotic dynamics can be made arbitrarily large
by increasing the time-delay and hence find applications in
secure communications. Most of the chaos based communication
techniques use synchronization in unidirectional drive response
system. A limitation which arises in this case is that messages
can be sent only in one direction. Thus for a two way
transmission of signals, a bidirectional coupling is required. In
this work we deal with an array of three mutually coupled
Josephson junctions with a finite time-delay in coupling  and study
the system dynamics in the presence of an external driving
field. The effect of phase difference and  a small frequency
mismatch  between the driving fields on a chaotically
synchronized system with time delay is then studied.}
\section{Introduction}
The equation for Josephson junction (JJ) when treated within the
Stewart-McCumber model is identical to the equation for a driven
damped pendulum which has been studied theoretically for several
routes to chaos\cite{humier,braiman}. Since the work done by
Belykh, Pedersen and Soerensen on chaotic dynamics in Josephson
junction (JJ) \cite{bel,bely}, a great deal of investigations
have been carried out as it is an interesting candidate in the
field of non linear dynamics. A detailed review of chaos in JJ
can be found in \cite{kau}.
 Time-delay has been studied experimentally
in JJ transmission lines represented by a sequence of overdamped
underbiased JJs \cite{kapl}. The synchronous variation of all
the bias currents causes a change in the time required for the
pulses to pass through the line, and hence possible to provide a
required time delay.

A generalized stability theory for synchronized motion of
coupled oscillator systems was developed by Fujisaka and Yamada
\cite{fuji,fujis}. Since Pecora and Carroll \cite{pecora}
reported the synchronization of chaotic systems, different types
of synchronization such as complete, generalized and phase
synchronization of chaotic oscillators have been described
theoretically and observed experimentally in different physical
situations\cite{pik,kur}. Intermittent synchronization has been
reported in bidirectionally coupled chaotic Josephson junction
\cite{bla}. Chaos synchronization in coupled JJ using active
control techniques has also been studied \cite{uca}. We have
studied the effect of phase difference between applied fields on
suppression of chaos and synchronization in parallely coupled
JJs \cite{chi}. Time-delay is ubiquitous in physical, chemical
and biological systems due to finite transmission times,
switching speeds and memory effects.  Anticipatory, lag,
projective  and phase synchronizations have been reported in
time-delayed systems \cite{fen,feng,sen}. However the addition
of a time-delay considerably complicates stability analysis. The
condition for stability  of synchronization for unidirectionally
coupled piece-wise linear time-delayed system can be obtained
using the Lyapunov-Krasovskii method \cite{kra,pyr}.
Intermittent anticipatory, intermittent lag and complete
synchronization were found to exist at the same time in
unidirectionally coupled nonlinear time-delayed system having
two different time-delays \cite{sen1}. Important functional
information about the brain states could be obtained by
modelling it with equivalent scale-free small-world networks
\cite{egu}. A configuration for such a model was given by
considering three bidirectionally coupled  oscillators in a
line. The outer elements were seen to synchronize isochronally
while the middle element lagged behind. It was also verified for
a neuron model with three elements that the outer elements gets
synchronized while remaining lag synchronized with the middle
one\cite{fis}. The phenomena of synchronization of the outer
lasers in a system of three lasers has been explained using the
ideas of generalized synchronization \cite{alex}. Isochronal
synchrony in coupled semiconductor and fiber ring laser models
with mutual delay-coupling was also studied \cite{ira}.

 In this work, we chose three mutually
coupled Josephson junctions with time-delay in coupling as a
model for our studies. To our knowledge,  mutually coupled
systems described by a  second order differential equation with
time-delay has not been studied earlier for phase effects on
synchronization. In section II, the model is introduced and
section III deals with the stability analysis. The effect of
phase difference and frequency detuning among the applied fields
on the system is discussed in section IV. In section V, we
discuss the results and present a conclusion.
\section{The Model}
The normalized equation of a single Josephson junction as
represented by the resistively and capacitively shunted (RCSJ)
model is \cite{bar}
\begin{equation}
\label{single} \ddot{\phi}+ \beta \dot{\phi}+
\sin\phi=i_{dc}+i_0 \cos(\Omega t),
\end{equation}
where $\phi$ is order parameter phase difference,
$\beta=\frac{1}{R}\sqrt{\frac{\hbar}{2 e i_{c} C}}$ is the
damping parameter,   $i_{dc}$ is the applied dc biasing and
$i_0\cos(\Omega t)$ is the applied rf-field. The dynamical
equations for three Josephson junctions with a time-delay in
coupling  can be given as
\begin{eqnarray}
\label{jjcoup} \nonumber \ddot{\phi_1}+ \beta \dot{\phi_1}+
\sin\phi_1&=&i_{dc} + i_0 \cos(\Omega t)
-\alpha_s\left[\dot{\phi_1}-\dot{\phi_2}(t-\tau)\right] \\
\nonumber \ddot{\phi_2}+ \beta \dot{\phi_2}+
\sin\phi_2&=&\alpha_s\left[\dot{\phi_1}(t-\tau)+ \dot{\phi_3}(t-\tau)-2 \dot{\phi_2}\right] \\
\ddot{\phi_3}+ \beta \dot{\phi_3}+
\sin\phi_3&=&i_{dc}+i_0\cos[(\Omega+\Delta\Omega)t+\theta]
-\alpha_s\left[\dot{\phi_3}-\dot{\phi_2}(t-\tau)\right],
\end{eqnarray}
where $\tau$ is the time delay applied and
$\alpha_s$ is the coupling term which takes into account of the
contribution from the delay circuit. $\Delta\Omega$ is the
frequency detuning  and $\theta$ is the phase difference among
the applied fields. Phase synchronization has been studied in a
similar system of chaotic rotators without time-delay
\cite{grig}. In the present work we have considered the systems
to be identical and Eq. (\ref{jjcoup}) may be written in
equivalent form as

\begin{eqnarray}\nonumber
\label{runge}
\dot{\phi_1} &=& \psi_1\\\nonumber %
\dot{\psi_1}&=& -\beta \psi_1-\sin\phi_1+i_{dc}+i_0 \cos(\Omega
t)
-\alpha_s \left[\psi_1-\psi_2(t-\tau)\right]\\
\dot{\phi_2} &=& \psi_2\\\nonumber %
\dot{\psi_2}&=& -\beta \psi_2-\sin\phi_2
+\alpha_s\left[\psi_1 (t-\tau)+ \psi_3 (t-\tau)-2 \psi_2\right]\\\nonumber
\dot{\phi_3} &=& \psi_3\\\nonumber %
\dot{\psi_3}&=&-\beta \psi_3-\sin\phi_3+i_{dc}+i_0
\cos[((\Omega+\Delta\Omega) t+\theta)]-\alpha_s
\left[\psi_3-\psi_2(t-\tau)\right].
\end{eqnarray}
For the case $\Delta\Omega=\theta=0$,   it can be seen from Eq.
\ref{jjcoup} that the subsystem consisting of the outer JJs
possess symmetry with respect to interchange of variables and
hence may possess identical solutions. This type of situations where
identical solutions exists for coupled systems is known as complete synchronization.
Chaotic systems also exhibit complete synchronization for some values of parameters.
An array of JJs with no delay in coupling is found to be chaotically
synchronized for the parameter values $\beta=0.3, i_0=1.2,
\Omega=0.6, i_{dc}=0.3$ with $\alpha_s=0.37$ \cite{chitra}.
We have selected these values of parameters for numerical studies
unless specified otherwise. In the presence of a time-delay in
coupling, it is found that the dynamics  changes between
periodic and chaotic motions. However synchronization is found
to be unaffected by the time delay.

\section{Stability analysis}
In order to  check for the stability of  synchronization  and
its dependence on various parameters, we need to know the
transverse Lyapunov exponents (TLE) and its dependence on the
parameters. The necessary and sufficient condition for stability of the synchronous
solution is that all the transverse Lyapunov exponents (TLE)
calculated with respect to the perturbation out of the
synchronization manifold should be negative. However
calculation of Lyapunov exponents gets complicated
when time-delays are involved. By linearizing the equation for
the outer JJs about the synchronization manifold, we can  arrive at a
necessary condition for synchronization, i.e., the sum of the
TLE should be negative. If the sum of the TLEs is negative, it
implies that the phase space is shrinking. Let $\phi(t)$ and
$\psi(t)$ represent the synchronous solution and we define new variables $\Delta\phi_i(t)=\phi_i(t)-\phi(t)$ and
$\Delta\psi_i(t)=\psi_i(t)-\psi(t)$ with $i=1,3$. Here
$\Delta\phi_i(t)$  and $\Delta\psi_i(t)$ are the perturbations
of the outer oscillators from the synchronization manifold.
Linearizing Eq.\ref{runge} transverse to the synchronization
manifold , we get after dropping the subscripts \cite{alex}
\[ \left( \begin{array}{c}\Delta\dot{\phi}\\ \Delta\dot{\psi}\\ \end{array} \right) =
\left( \begin{array}{cccc}
0 &  &  & 1 \\
1 &   &  & -\beta-\alpha_s \\
 \end{array} \right) \left( \begin{array}{c}\Delta\phi\\ \Delta\psi\\ \end{array}
 \right).\]\
We have approximated $\sin(\Delta \phi)\approx \Delta \phi$ as
$\Delta \phi$ is small. Due to the delay in coupling,
perturbations will not affect the coefficient matrix until
$t-t_1\geq 2 \tau.$  The Wronskian of the linearized system can
be related to the trace of the
 matrix by Abel's formula
 \[ W(t) = \left| \begin{array}{cc}
\Delta\phi & \Delta\psi  \\
\Delta\dot{\phi} & \Delta\dot{\psi}  \\
 \end{array} \right|=exp\left(\int_{t_1}^{t}(-\alpha_s-\beta)dt\right).\]
The Wronskian gives the phase space dynamics of the system. Taking
the natural log of the Wronskian we get
\begin{equation}
\label{logar}
\ln[W(t)]=\ln|\Delta\phi\Delta\dot{\psi}-\Delta\psi\Delta\dot{\phi}|=-\int_{t1}^t(\alpha_s+\beta)
dt.
\end{equation}
This is a monotonically decreasing function of $t$ which means
that the phase space volume of the system perturbed from the
synchronization manifold contracts as a function of time. The
sum of the transverse Lyapunov exponents is given as
\begin{equation}
\label{lyapunov} \sum_{j=1}^{2}\lambda_j=\lim_{t\rightarrow
\infty} \frac{1}{t}ln|\Delta\phi\Delta\dot{\psi}-\Delta\psi\Delta\dot{\phi}|,
\end{equation}
 The sum of the transverse Lyapunov exponents can
be now approximated using Eq. \ref{logar} as
\begin{equation}
\label{sumlya}
\lambda_1+\lambda_2\approx-(\alpha_s+\beta).
\end{equation}
which is negative indicating that the phase space of the coupled
system shrinks to a trajectory representing the synchronous
state. The  sum of the Lyapunov exponents depends on the
coupling term and damping parameter.  Even though the  sum of
the conditional Lyapunov exponents is negative, if one of the
exponents is positive, the solution will blow up along the
unstable direction. This kind of a situation, where isochronally
synchronized solution gets unstable  even though the sum of
Lyapunov exponents is negative was addressed in ref.~\cite{ira}.

\begin{figure}[h]
\centering
\includegraphics[width=8cm]{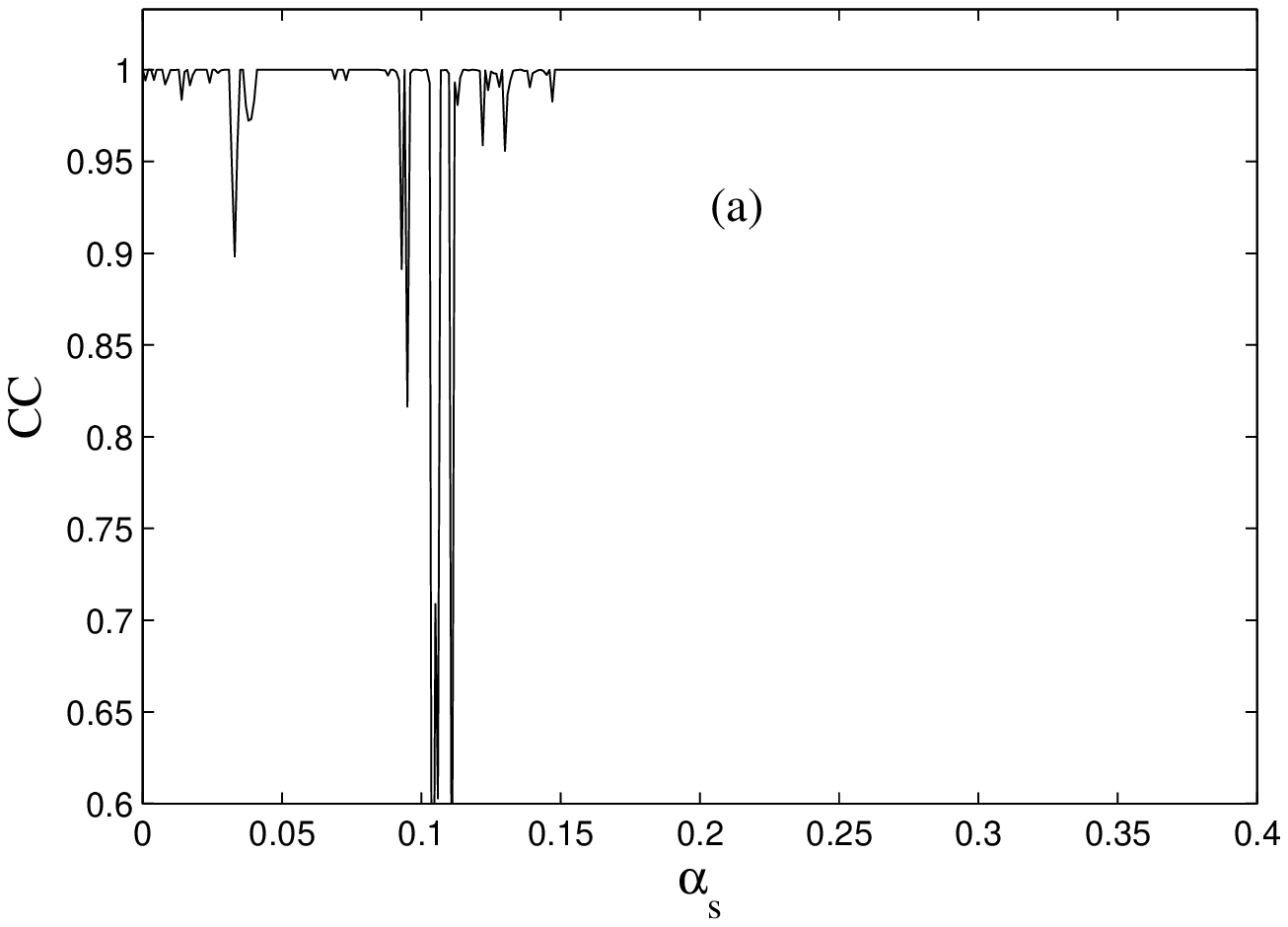}%
\hspace{1cm}%
\includegraphics[width=8cm]{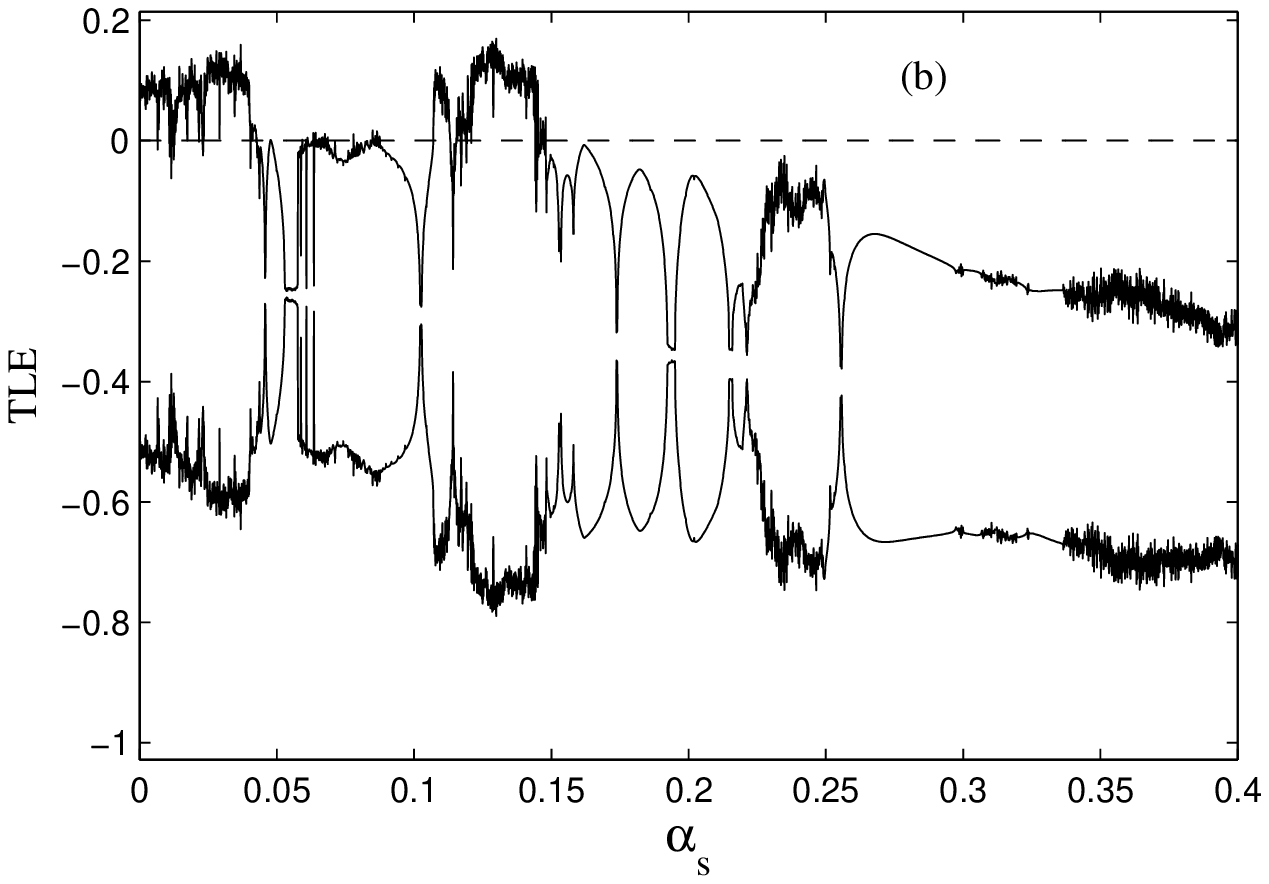}
\caption{(a)Crosscorrelation (CC) between the two outer JJs for
different values of coupling strength with all other values
$\beta=0.3, i_0=1.2, \Omega=0.6, i_{dc}=0.3$ (b) Transverse
Lyapunov exponents plotted for the same parameter values.
$\tau$=0.1 in both cases.
} %
\label{crosscorr}
\end{figure}

\begin{figure}[th]
\centering
\includegraphics[width=8cm]{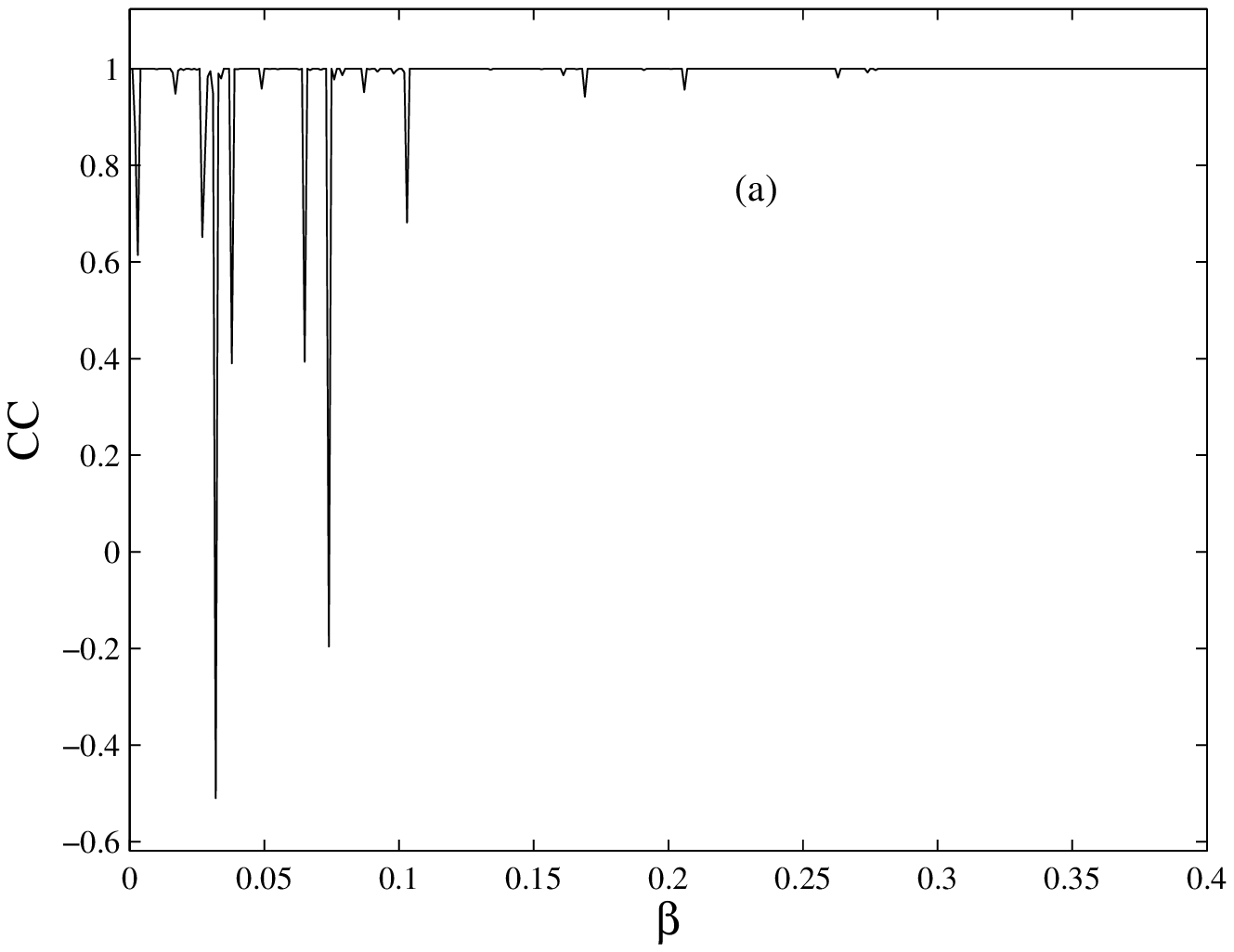}
\hspace{1cm}%
\includegraphics[width=8cm]{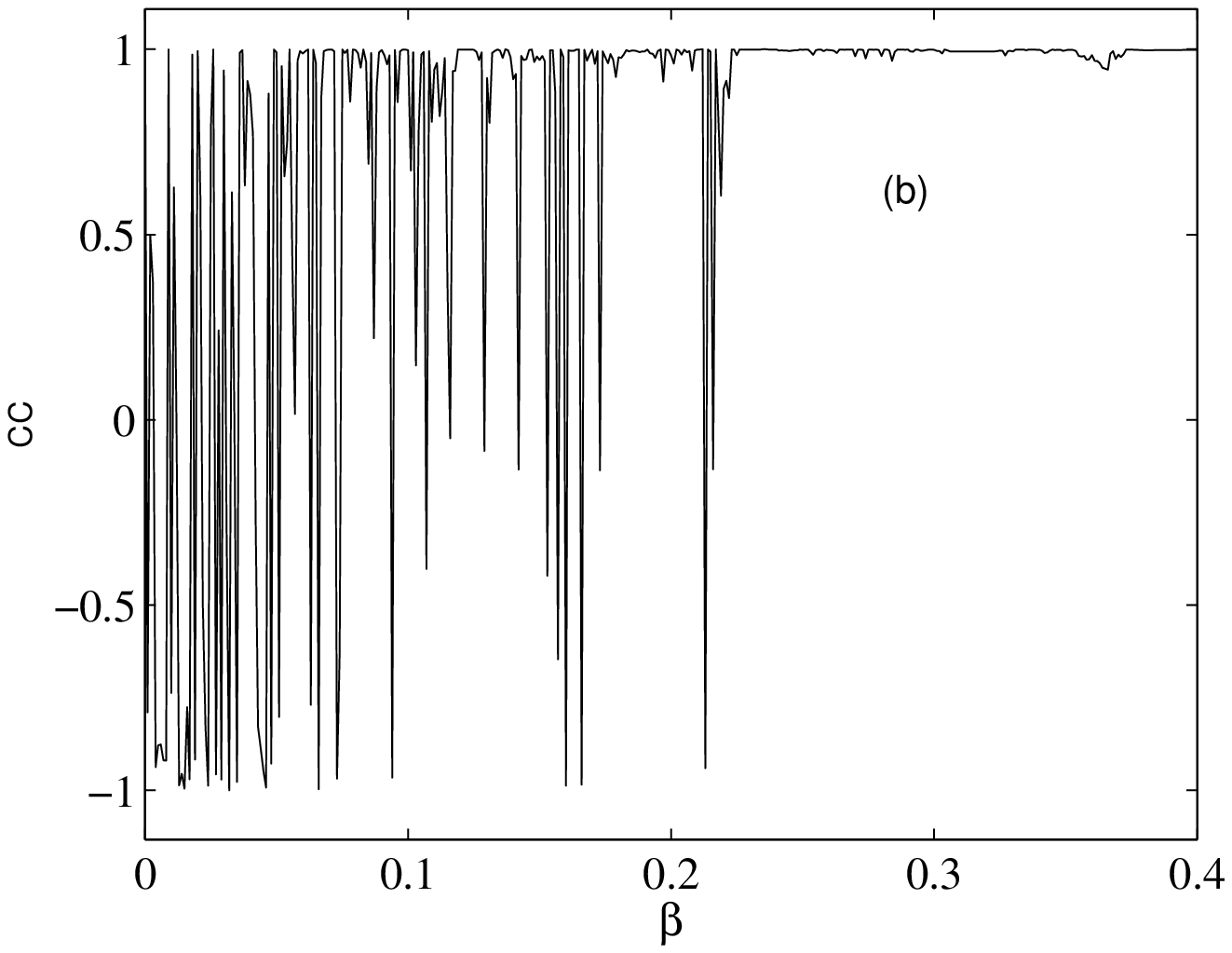}
\caption{(a) CC between the two outer JJs for different values
of damping parameter with $\alpha_s=0.37$ (b)CC between the
outer and middle junction.} %
\label{corrbet}
\end{figure}

 The quality of synchronization is usually quantified using the correlation coefficient (CC) given by
\begin{equation}
 CC=\frac{\langle \left[ \psi_1(t)-\langle \psi_1(t) \rangle \right]\left[ \psi_i(t)-\langle \psi_i(t) \rangle \right] \rangle}{\sqrt{\langle \left\vert \psi_1(t)-\langle \psi_1(t) \rangle \right\vert^2 \rangle}\sqrt{\langle \left\vert \psi_i(t)-\langle \psi_i(t) \rangle \right\vert^2 \rangle}}
\end{equation}
with $i=2,3.$ The value of correlation coefficient lies between
$-1\leq CC \leq 1$ with  large value of $\mid CC \mid$ meaning
better synchrony. The cross correlation of the dynamics is shown
in Fig.\ref{crosscorr}(a) for the JJ system, from which we
observe that for very low values of coupling constant there is
loss of synchrony. In Fig.\ref{crosscorr}(b) the transverse
Lyapunov exponents are plotted. It can be seen that the sum of
the TLE will always be negative as given by Eq.\ref{sumlya}.We
observe that corresponding to the values where cross correlation
is lost, there is a small positive value for one of the TLE.
Fig.\ref{corrbet}~(a) shows the cross correlation between the
outer junctions for various values of damping parameter. It can
be observed from Fig.\ref{corrbet}~(b) that the middle junction
remains uncorrelated with the outer ones for most of the values
of damping parameter. Thus the middle junction mediates
synchronization between the outer junction while  remaining
unsynchronized from both the outer junctions.

\begin{figure}[th]
\centering
\includegraphics[width=8cm]{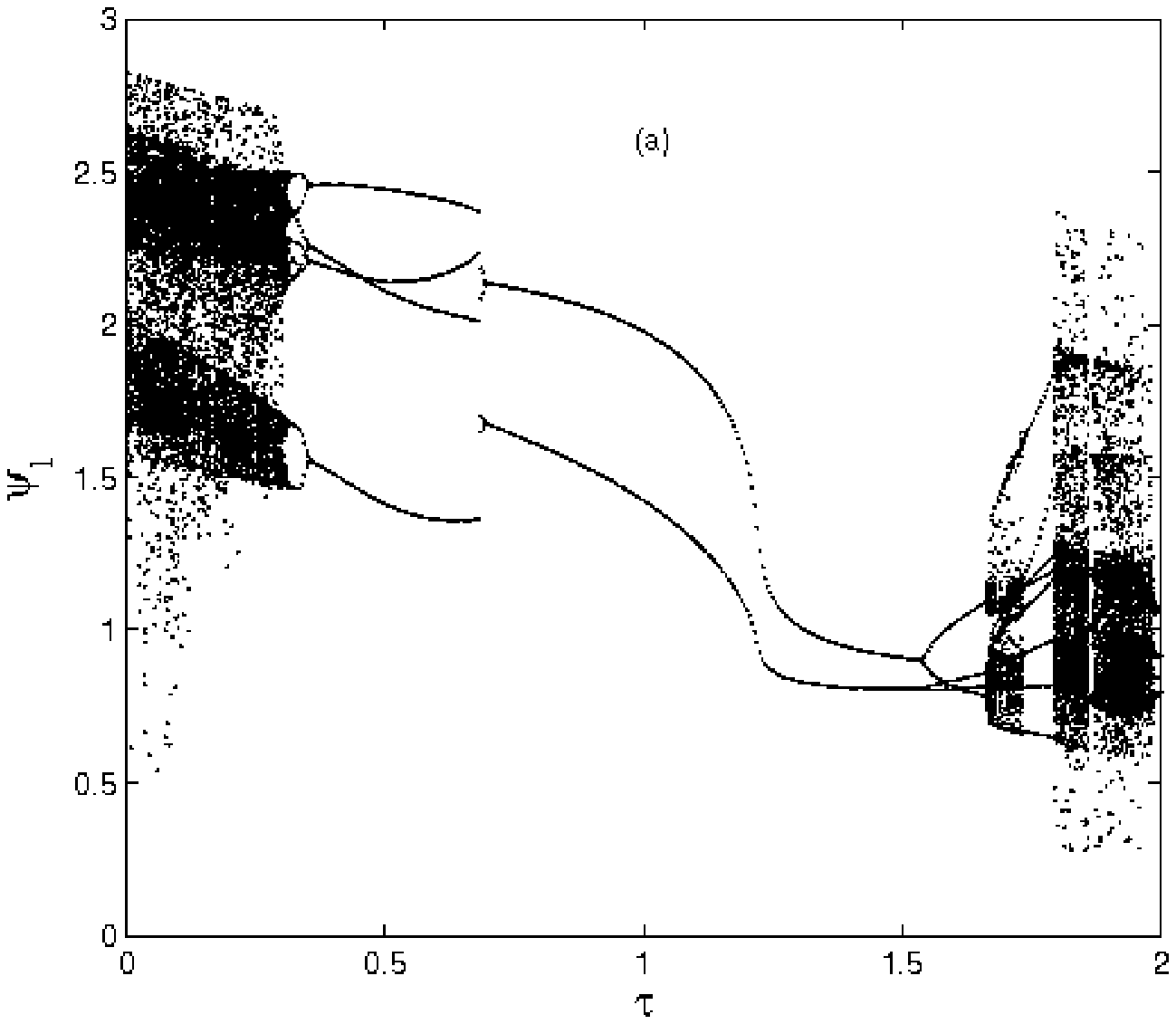}%
\hspace{1cm}%
\includegraphics[width=8cm]{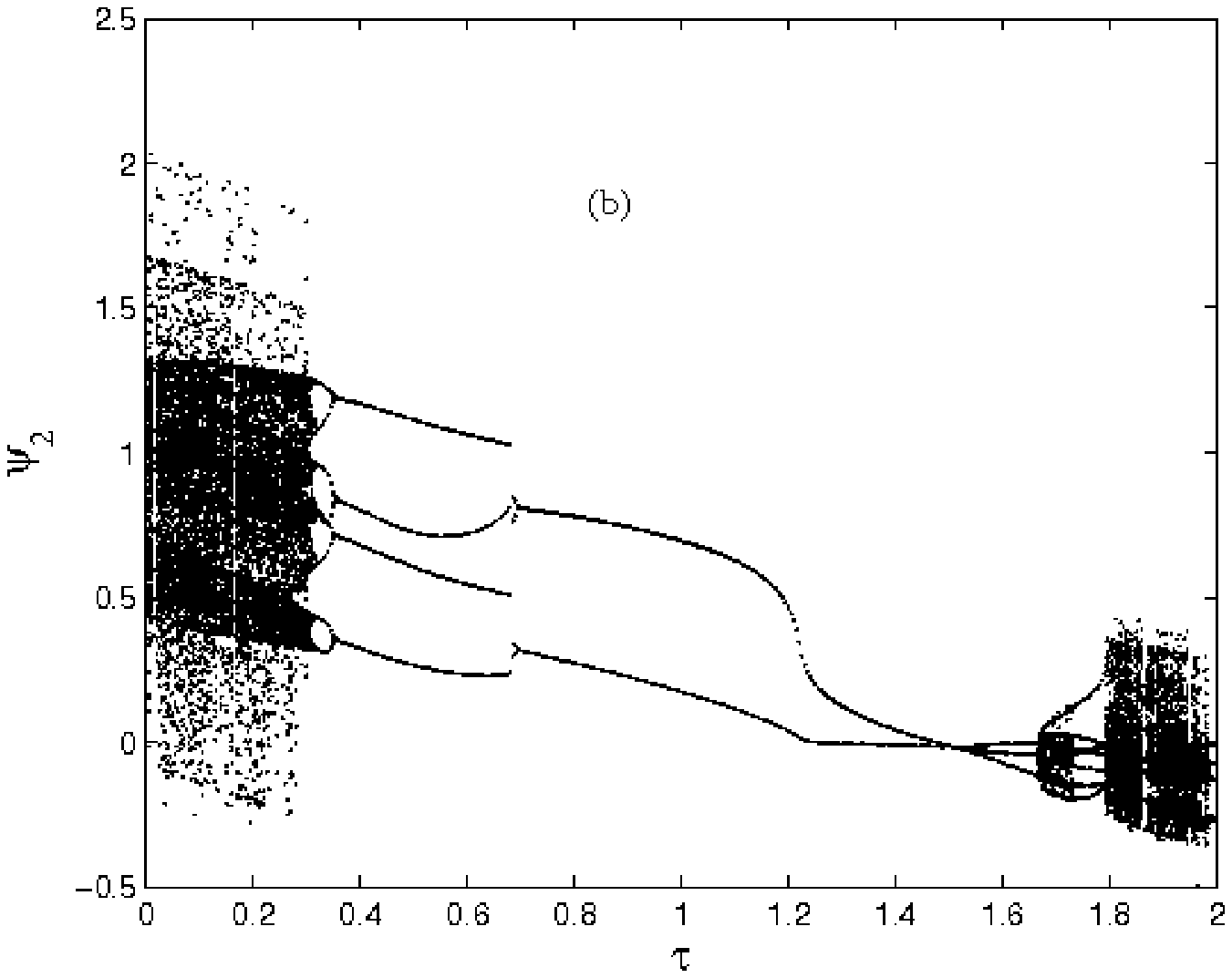}
\caption{(a)Bifurcation diagram for $\psi_1$ against $\tau$
(b)Bifurcation diagram for $\psi_2$ against $\tau$. Parameter values are $\beta=0.3,i_0=1.2,\Omega=0.6,i_{dc}=0.3,\alpha_s=0.37$} %
\label{powtau7}
\end{figure}

\begin{figure}[th]
\centering
\includegraphics[width=8cm]{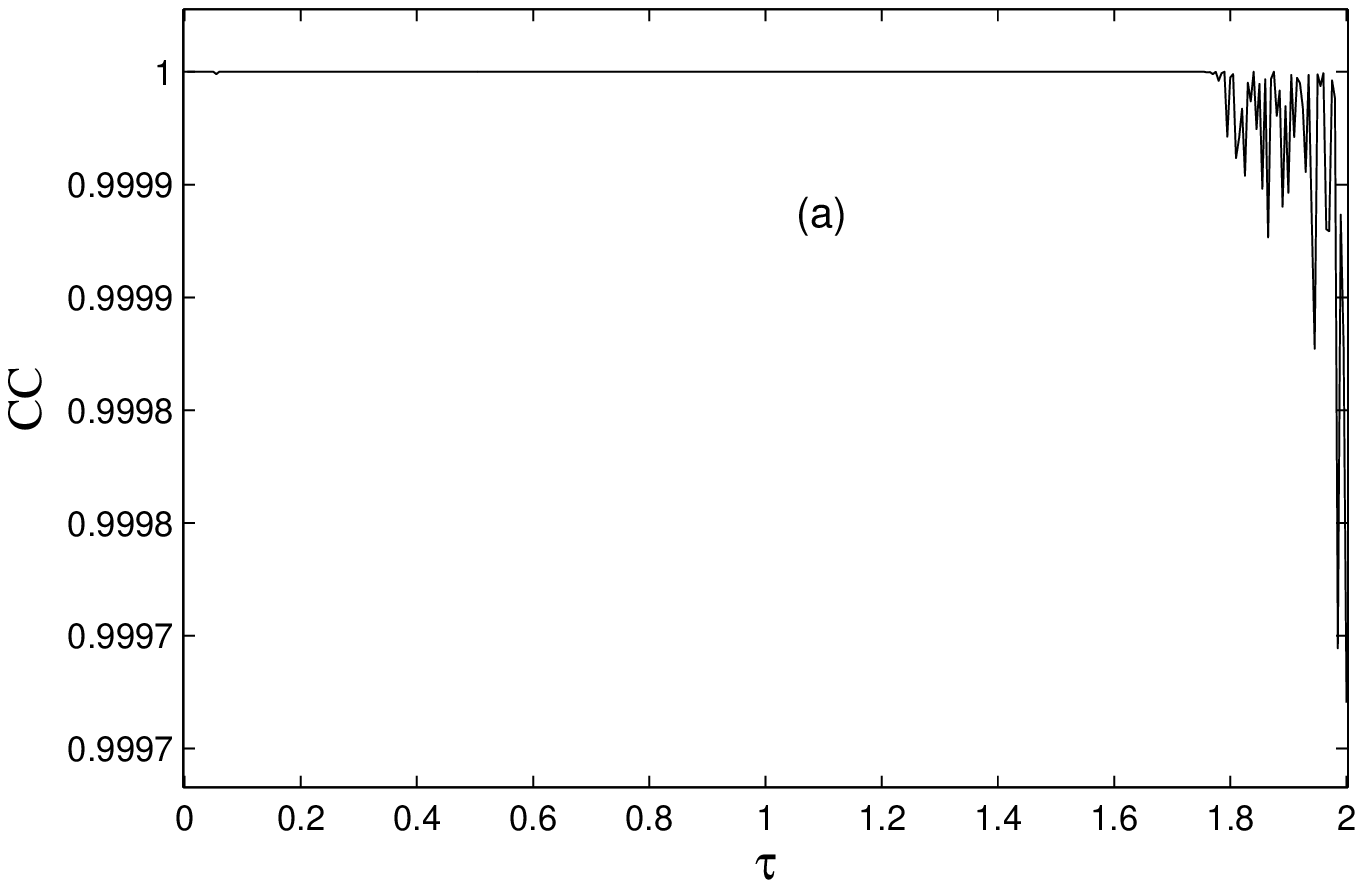}
\hspace{1cm}%
\includegraphics[width=8cm]{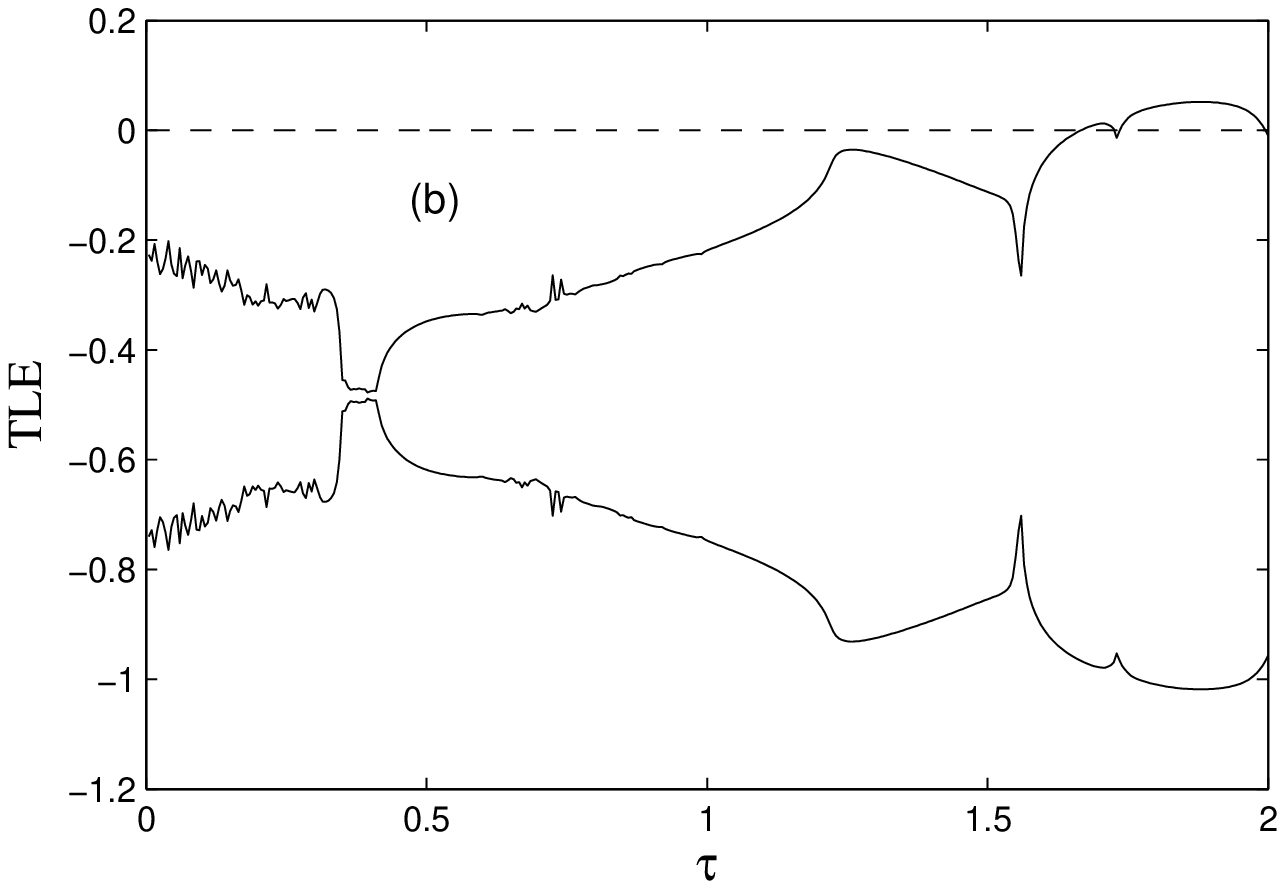}
\caption{(a)Cross correlation for various values of $\tau$ with
the other parameter values as $\beta=0.3, i_0=1.2,
\Omega=0.6, i_{dc}=0.3, \alpha_s=0.37$ (b) TLEs  plotted for different values of $\tau$} %
\label{corrtau}
\end{figure}

In the presence of a time-delay $\tau$, the dynamics of the
system changes considerably. It is observed that for some values
of time delay (for $\tau$ nearly equal to $0.35$ onwards) the
system exhibits periodic synchronized motion.  Fig.
\ref{powtau7}(a) is the bifurcation plot for various time-delays
which shows periodic and chaotic behavior for different delay
times. With Fig. \ref{corrtau}(a), we show that the system
remains synchronized for most of the values of time-delay. The
TLEs plotted in Fig. \ref{corrtau}(b)shows that one of the TLE
has positive value for regions where CC is lost.

\section{Effect of phase difference and frequency detuning}
A phase difference between the applied fields was found to supress chaos effectively
in an array of coupled Josephson junctions. In this section we study the effect of phase
difference in bidirectionally coupled time-delay systems. From Eq. \ref{jjcoup}, we can see
that in the presence of an applied phase difference, the equation for the outer junctions
is no longer identical. In terms of the difference variable $\psi_{1,3}^-=\psi_1-\psi_3$
we can write the equation for the outer junctions from Eq. \ref{runge} as
\begin{equation}
 \dot{\psi}_{1,3}^-= -\beta \psi_{1,3}^- -2 \cos(\frac{\phi_{1,3}^+}{2}) \sin(\frac{\phi_{1,3}^-}{2})+i_0' \sin(\Omega t + \theta/2)-\alpha_s\psi_{1,3}^-
\end{equation}
where $\phi_{1,3}^-=\phi_1-\phi_3$, $\phi_{1,3}^+=\phi_1+\phi_3$ and $i_0'=2 i_0\sin(\theta/2)$
and it can be seen that for a finite value of phase difference the outer junctions cannot get synchronized.

\begin{figure}[ht]
\centering
\includegraphics[width=8cm]{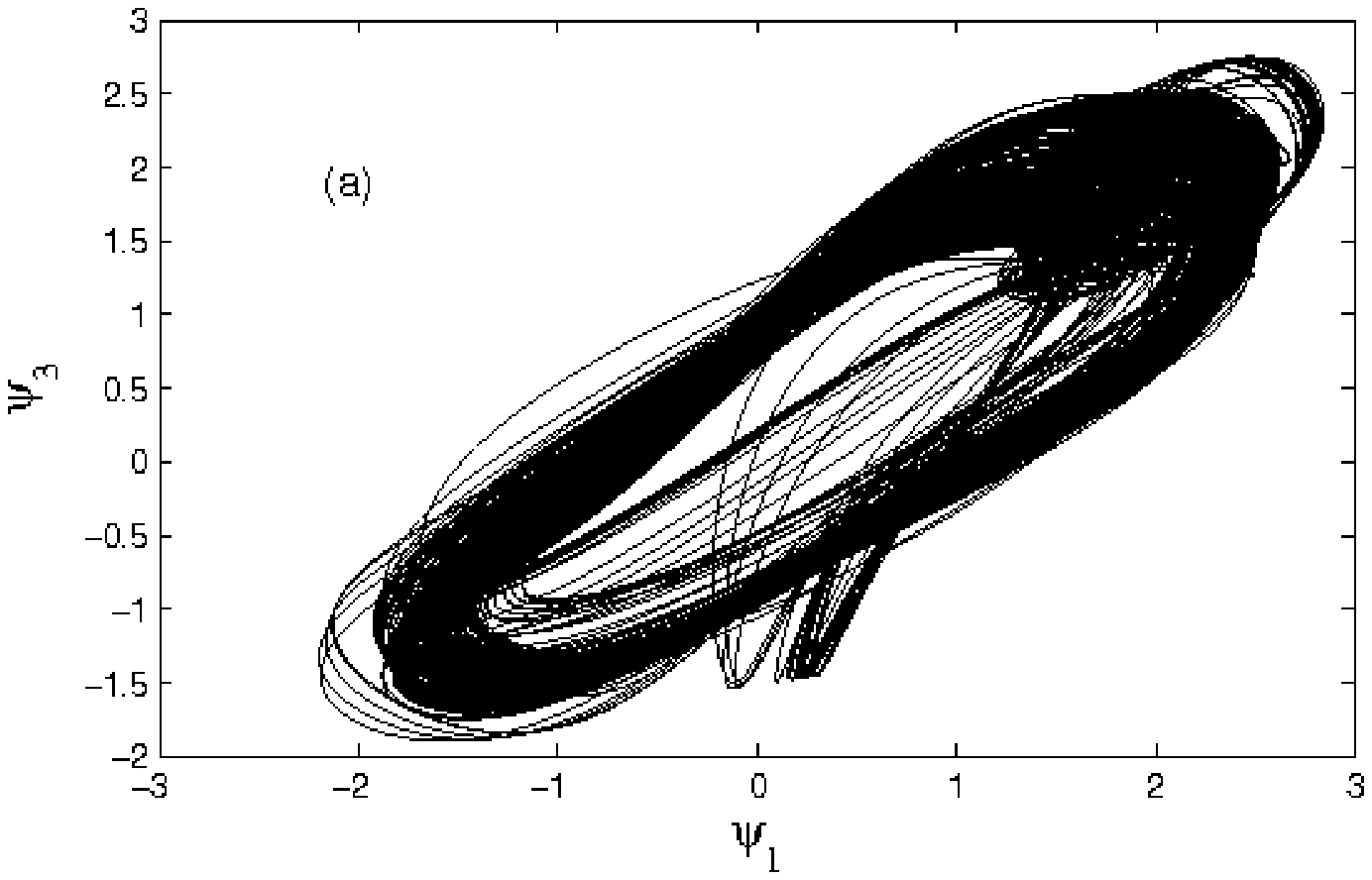}
\hspace{1cm}%
\includegraphics[width=7cm]{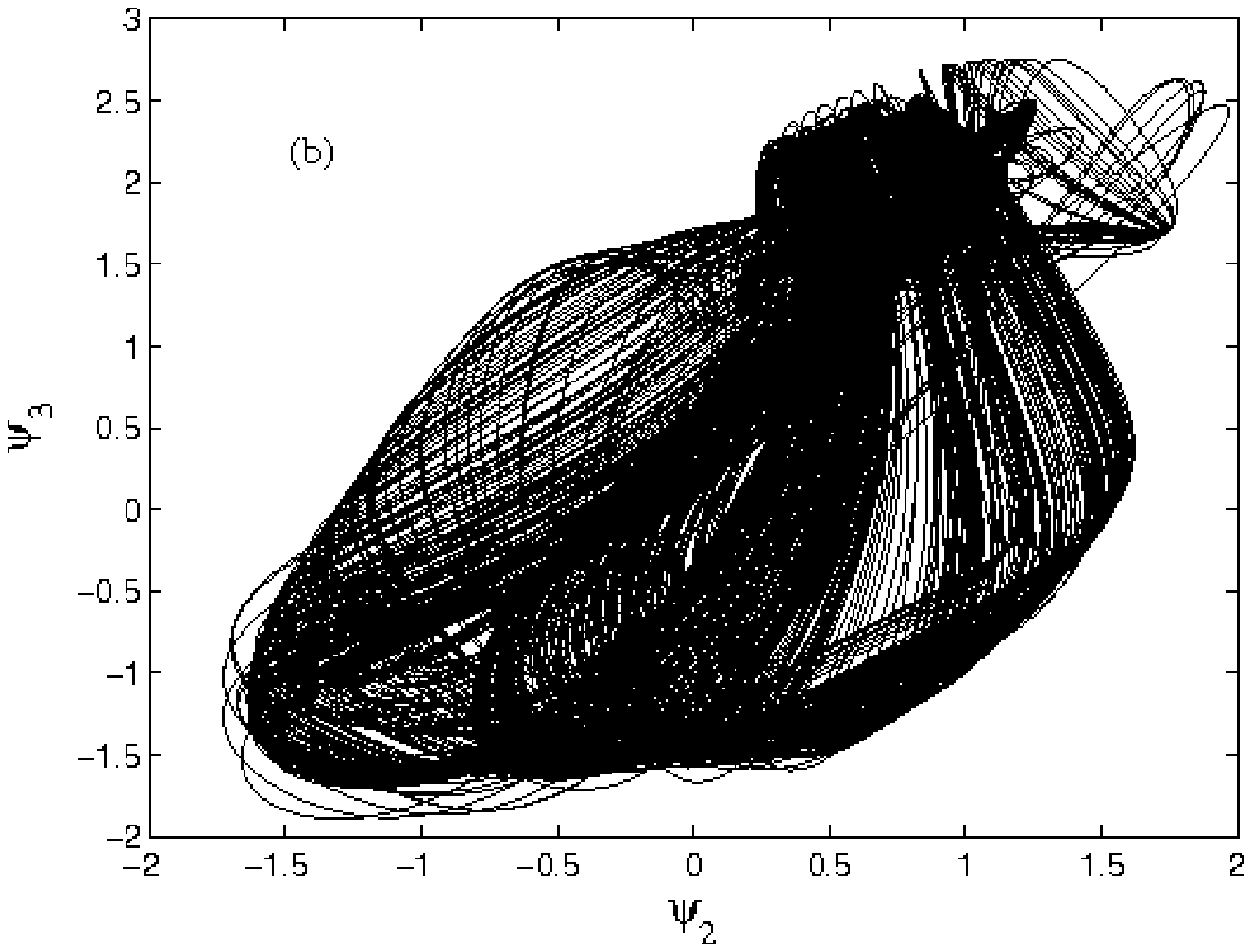}
\caption{The outer and inner junctions are desynchronized on the application of a phase difference $\theta=0.15 \pi$ with $\tau=0.$} %
\label{synthep1tau0}
\end{figure}

\begin{figure}[ht]
\centering
\includegraphics[width=8cm]{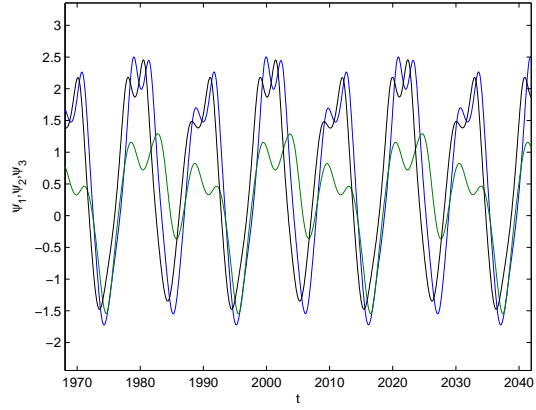}
\caption{Periodic when a time delay $\tau=0.25$ along with a phase difference $\theta=0.15 \pi$ applied.} %
\label{timthep1tau10}
\end{figure}

The phase difference which usually desynchronizes the system may be used to suppress chaos in time-delayed systems.
From Figs.~\ref{powtau7} and \ref{corrtau}, it can be seen that for a time delay of $\tau=0.25$ the system is  chaotic and synchronized.
Fig. \ref{synthep1tau0} shows that an applied
phase difference of $\theta= 0.15 \pi$ desynchronizes the system with $\tau=0$. But when a
time-delay along with a phase difference is applied the system exhibits periodic motion.
From the time series plotted in Fig.~\ref{timthep1tau10}, it can be observed that the
application of a phase difference $\theta=0.15\pi$ along with
the time-delay, the system goes to a periodic state.

\begin{figure}[th]
\centering
\includegraphics[width=8cm]{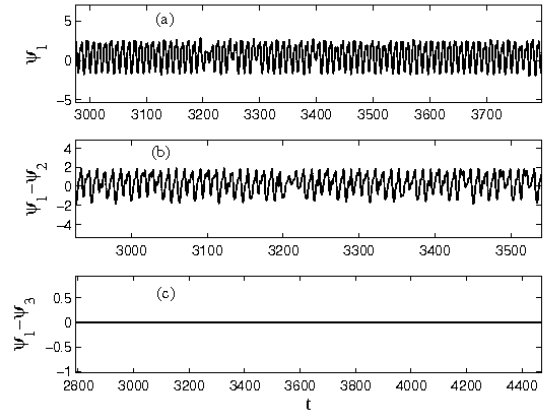}
\caption{The temporal dynamics of the variable with no detuning for $\alpha=0.37$} %
\label{detun00}
\end{figure}

To complete the discussion, we consider the case of frequency
detuning  as it is difficult to apply a
second frequency which is exactly  same as the  first. Small deviations
in the applied frequencies are inevitable. So we consider the
system with a small frequency detuning and the initial phase
difference is taken to be zero. In order to investigate the
difference in the dynamics brought about by frequency detuning,
we first check the case where $\Delta\Omega=0$ in Eq.\ref{jjcoup}, i.e, the case
without any detuning. Fig. \ref{detun00} shows the temporal
dynamics of the system considered with no detuning. Fig. \ref{detun00}(a) shows the time series plot for  $\psi_1$
with no detuning.  Fig. \ref{detun00}(b) shows the time series
plot for the difference between the  variables of the outer and the middle junction while Fig. \ref{detun00}(c) is time series plot for the outer junctions.
The outer junctions are perfectly synchronized while remaining
uncorrelated with the  inner  junction.

\begin{figure}[tbh]
\centering
\includegraphics[width=8cm]{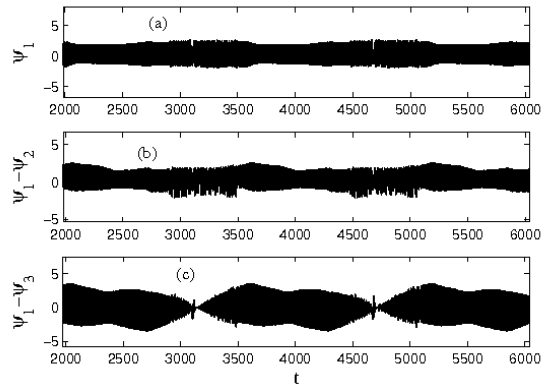}~
\caption{The temporal dynamics of the variable with $\Delta\Omega=0.004$ for $\alpha=0.37$} %
\label{detun003}
\end{figure}

When a frequency detuning is applied, it can be observed that
the outer junctions show synchronization in regular intervals. Fig.~\ref{detun003}(a)
is the time series plot for the variable for the outer junction
with $\Delta\Omega=0.004$ and it can be seen that a periodic
modulation has appeared due to detuning.  Here we observe that
though the qualitative behavior is repeated, the trajectory of
the system cannot be completely repeated due to the chaotic
segments in the evolution process. This type of motion is
referred to as breathers. Fig.~\ref{detun003}(b) shows the time series plot for the  difference between the variables of the outer and middle
junctions and it can be seen that they remain uncorrelated.  Fig.~\ref{detun003}(c) shows the time series plot for the difference between the variables of
 the outer junctions. It can be seen that they get synchronized
 periodically with a period of $T=2 \pi /\Delta\Omega $.
The qualitative behavior of the system is not affected by time
delays.

\section{Result and Discussion}
In this work we deal with bidirectionally coupled time-delayed
systems and study the effect of delay and phase difference
between the applied fields on synchronization. The sum of the
Lyapunov exponents transverse to the synchronization manifold is
evaluated and it is found to be negative indicating the
shrinking of phase space of the coupled system to a trajectory
representing the synchronous state. However cross correlation
coefficient reveals positive TLE for some values of coupling
constant and damping parameter where synchrony is lost, though
the sum would be still negative.  Transverse Lyapunov exponents
are evaluated numerically and are found to be in good agreement
with the analytic results. By varying the time-delay, we have
analyzed the behavior of the system  and it is observed that for
small values of time delays, periodic motion occur and as the
delay time is increased chaotic motion reappears. However the
system remained synchronized for most of the values of
time-delay and hence may find applications in secure
communication. The study on a configuration of three oscillators
in a line is of importance because of the current recognition
that it has resemblance with neuron models. Phase difference
between the applied fields together with the time delays may be
effectively used to suppress chaos. The practical situation
where a small frequency detuning will be present between the
applied fields is also studied. As the frequency detuning is
like a time dependent phase difference, periodic and chaotic
motions are repeated with a period of $T=2 \pi/\Delta\Omega$.
Time-delay does not change the effect of frequency detuning.
Experimentally it is possible to apply a required delay in JJ
\cite{kapl}. Hence by suitably adjusting the time-delay, chaos
may be controlled in JJ devices like voltage standards,
detectors, SQUIDS etc where chaotic instabilities are least
desired.

\acknowledgements The authors acknowledge DRDO, Government of
India for financial assistance through a major research project.
The authors are grateful to Dr.Ira B. Schwartz, Naval Research
Laboratory, Nonlinear Systems Dynamics Sections, USA for the
valuable suggestions and also for the help extended during  the
numerical works. The authors thank the Reviewers for the
valuable comments which helped us to improve the original
manuscript.

\end{document}